# Impurity states of electrons on the surface of a nanotube in a magnetic field

G.I. Rashba
Department of Theoretical Physics named after I.M. Lifshits
Kharkiv National University named after V.N. Karazin
4 Svobody Sq., Kharkiv, 61022, Ukraine

e-mail: georgiy.i.rashba@gmail.com

**Abstract**

The localization of electrons in the field of isolated nonmagnetic impurity atoms on the surface of a nanotube in a magnetic field is considered. A model of a Gaussian separable impurity potential capable of localizing an electron at any intensity is used. The positions of the local level are found in the regime of strong and weak localizations. The positions of the resonances are found, their widths are estimated. They experience Aharonov-Bohm oscillations when the magnetic flux in the tube changes. It is shown that the positions of the resonances experience oscillations similar to the de Haas-van Alphen oscillations, which are not associated with a magnetic field.

## 1. Introduction

The development of modern nanoelectronics is impossible without the creation of theoretical methods for calculating the electronic properties of nanostructures [1, 2]. The presence of a stable atomic structure and unique physical properties causes considerable interest of researchers in single-walled carbon and semiconducting nanotubes. The unusual electrical properties of nanotubes make them one of the main materials in nanoelectronics [1, 2]. On the basis of nanotubes, electronic devices of nanometer (molecular) size are created. It is expected that in the foreseeable future nanotubes will replace elements of a similar purpose in electronic circuits of various devices, including modern computers. However, despite the fact that carbon and semiconductor nanotubes are known for the perfection of their structure, they may contain atomic-scale defects: impurities, vacancies, topological defects [2]. The presence of defects in nanotubes can be beneficial in achieving the desired functionality. It is well known that even one single structural defect can significantly change the electrical characteristics of such a one-dimensional conductor [3]. Therefore, information about how and to what extent various defects can change the electronic properties of nanotubes is very important. The point is that in this way a controlled engineering of the physical properties of nanotubes can be realized. In the future, this can lead to the emergence of various classes of devices with physical properties that are completely controlled by the creation of various defects.

Many years ago I.M. Lifshitz [4-10] created a new method in the theory of solids with impurity atoms, which is now called the method of local perturbations. This method is an integral part of a new direction in solid state theory – the theory of disordered systems, created mainly by I.M. Lifshitz and his collaborators. This method was used to calculate the characteristics of local and quasilocal (resonant, virtual) states of quasiparticles in the field of isolated impurity atoms in metals, semiconductors, dielectrics [11] and magnetics [12].

Impurity states of electrons in metals have been studied in many papers [13-23] and monographs [24-28]. The localization of electrons on impurity atoms in massive conductors and liquid metals was studied in Refs. [29, 30]. In Ref. [31], the theory of electron scattering by the $\delta$-potential of impurity atoms on the surface of a nanotube was developed. As an application of this theory, static conductivity is considered.

In the papers [32-34] the method by I.M. Lifshitz was used to study impurity states of electrons in three-dimensional conductors in a magnetic field. Currently the method of local perturbations is common in the physics of nanosystems. In papers [35-38] this method is used to consider impurity states of electrons in a two-dimensional electron gas. In papers [39–41], the method of local I.M. Lifshitz perturbations is employed to study the impurity states of electrons in quantum dots, quantum wires and on the surface of a nanotube. The physical properties of semiconductor nanotube in the metal conduction mode were discussed in connection with the problems of persistent currents in Ref. [42]. In this paper the persistent current was calculated taking into account the scattering of electrons by the $\delta$-potential of the impurity atom. The high-frequency properties of nanotubes have been discussed in Refs. [43-45].

Resonant states of electrons in metal carbon nanotubes in the field of impurity nitrogen atoms were found in Ref. [46]. The theory of these states in metallic and semiconductor nanotubes based on isolated impurities in the absence of a magnetic field is presented in Ref. [47].

Localization of electrons by the attractive $\delta$-potential on the surface of a cylindrical nanotube and in other systems in a magnetic field is considered in Ref. [48]. The authors of this paper regularized the diverging integral over the electron momentum in the dispersion equation for impurity levels of the electron energy by the cutoff method. In the process of renormalizing the theory, they expressed the cutoff momentum in terms of the "bare" coupling constant of an electron with a scattering center and its binding energy in the plane. Within the framework of the $\delta$-potential model, the authors of Ref. [48] discovered Aharonov-Bohm oscillations of the electron binding energy at an impurity center, studied the effect of magnetic flux in nanotube and the curvature of the structure on localization, and discovered unusual order-disorder phase transitions in an electron gas, induced by the flux [49].

In papers [33, 36-41] and in this paper another method of regularization of the integral in the dispersion equation for impurity levels was used. The "smeared" short-range potential of a nonmagnetic impurity atom in the form of a separable potential $|u\rangle u_0 \langle u|$, proposed by I.M. Lifshitz, is employed. Here $u_0$ is the coupling constant of an electron with an impurity center, $\langle \vec{r} | u \rangle$ is a certain function. It is chosen so, that the potential is as close as possible to the real one and is convenient for calculations. In papers [33, 36-41] the function $\langle \vec{r} | u \rangle$ is taken in the form of a Gaussian distribution.

In the Ref. [41] and in this paper, the "jellium" model, which is often applied in the solid state theory, is used. Zone effects are not considered. Their influence is manifested in replacing the mass of a free electron with an effective mass only. The dispersion law of the electron on the impurity-free nanotube can be taken as quadratic. Therefore, the results obtained here relate themselves mainly to semiconductor and carbon nanotubes in the metal conduction mode. The Ref. [41] considers an impurity on the surface of a nanotube in the form of the alien ring uniaxial to the tube axis. Such an impurity breaks the translational symmetry of the field in which the electron moves, while the axial symmetry remains unbroken. In Ref. [41] the authors limited themselves to the analysis of local states of electrons in the field of a ring attracting electrons and resonance states in a repulsive field. Resonant states in an attractive field were not considered in Ref. [41]. The Aharonov-Bohm oscillations induced by the magnetic flux, as well as the oscillations of the positions of the resonance levels, in case of changed tube parameters, not related to the magnetic field, were not discussed.

In the proposed paper, taking into account the magnetic field parallel to the nanotube axis, a short-range impurity potential is used, which is "smeared" not only along the nanotube axis, but also in the azimuthal direction. In Section 2, a model of the impurity potential is proposed and a dispersion equation for the impurity energy levels of electrons on the tube is obtained. In the Section 3 the binding energy of an electron in the field of such a potential is calculated. The Aharonov-Bohm oscillations of the electron binding energy are considered. Also in this Section, the positions of resonances in the attractive field, their damping are calculated, and oscillations not related to the magnetic field are found. Section 4 presents the main results of the paper. Section 5 contains the interpretation of the results obtained in the paper and their comparison with the already known results. In the Conclusion, the results of the paper are briefly summarized.

## 2. Theoretical Basis

### 2.1. Model of impurity potential on a tube

The function mentioned above $\langle \vec{r} | u \rangle$ on a tube surface is chosen as

$$\langle \varphi, z | u(\varphi, z) \rangle = u(\varphi) u(z),$$

where

$$u(\varphi) = \left( \frac{2}{\pi \eta} \right)^{1/4} \exp\left( -\frac{\varphi^2}{\eta} \right),$$

$$u(z) = \left( \frac{2}{\pi} \right)^{1/4} \frac{1}{\sqrt{b}} \exp\left( -\frac{z^2}{b^2} \right).$$

(1)

Here $\varphi, z$ are the cylindrical coordinates, $\eta$ and $b$ are the constants introduced to regularize the integral over the electron momentum in the dispersion equation. They characterize the length of the potential in the azimuthal direction and along the axis of the nanotube. In the limit $\eta \to 0, b \to 0$ the functions (1) are transformed into $\delta(\varphi)$ and $\delta(z)$. They are normalized by the conditions $\langle u(\varphi) | u(\varphi) \rangle = 1, \langle u(z) | u(z) \rangle = 1$. The projections of functions (1) onto the basis states of electrons $e^{im\varphi} \cdot e^{ikz}$ are

$$u_m = (2\pi\eta)^{1/4} \exp\left(-\frac{m^2\eta}{4}\right),$$
$$u_k = (2\pi)^{1/4} \sqrt{b} \exp\left(-\frac{k^2 b^2}{4}\right), \qquad (2)$$

where $m$ and $k$ are the projections of the angular momentum and momentum of the electron onto the axis of the nanotube. The quantum constant is chosen equal to unity. The amplitude of electron scattering by an impurity potential in the separable representation contains

$$(u_m u_k)^2 = \sqrt{\eta} 2\pi b \exp\left(-\frac{m^2\eta}{2}\right) \cdot \exp\left(-\frac{k^2 b^2}{2}\right). \qquad (3)$$

## 2.2. Dispersion equation for impurity energy levels

The scattering amplitude of electrons T by an impurity atom is related to the electron propagator $G$ by the relation $G = G_0 + G_0 T G_0$, where $G_0$ is the propagator of a free electron on the tube. The amplitude is calculated using the multiple scattering theory. This means taking into account the diagrams for the amplitude with one cross in terms of the cross technique [50].

In a separable representation, the amplitude has the form

$$T_\sigma(\varepsilon) = u_0 \sum_{mk} (u_m u_k)^2 G_\sigma^0(m,k,\varepsilon), \qquad (4)$$

where

$$G_\sigma^0(m,k,\varepsilon) = \frac{1}{\varepsilon - \varepsilon_{mk\sigma} + i0}$$

is the Green's function of free electrons in $(m,k)$-representation, $\sigma = \pm 1$ is the spin quantum number, $\varepsilon_{mk\sigma}$ is the energy of an electron [41] with an effective mass $m_*$ in the state $|mk\sigma\rangle$ equals

$$\varepsilon_{mk\sigma} = \varepsilon_0 \left(m + \frac{\Phi}{\Phi_0}\right)^2 + \frac{k^2}{2m_*} + \sigma\mu B, \qquad (5)$$

где $\mu$ is the Bohr magneton, $\varepsilon_0 = 1/2m_* a^2$ is the rotational quantum, $a$ is the nanotube radius, $\Phi$ is the magnetic induction $B$ flux in a nanotube, $\Phi_0 = 2\pi c/e$ is the flux quantum [51]. Energy (5) does not change during transformation $m \to -m, \sigma \to -\sigma, k \to -k, B \to -B$. In Eq. (4), the amplitude of electron scattering by an isolated impurity atom is taken into account exactly. In the realistic case of a many-impurity problem on a tube, as a result of averaging the electron propagator over the positions of impurity atoms, only terms linear in the concentration of scattering centers are taken into account.

The poles of the scattering amplitude (4) give the positions and widths of the impurity energy levels of electrons. Writing down the denominator in Eq. (4) in the form

$$1 - \frac{u_0}{4\pi^2 a}\left[F_\sigma(\varepsilon) - i\pi g_\sigma(\varepsilon)\right], \qquad (6)$$

where

$$F_\sigma(\varepsilon) = \sum_{m=-\infty}^{\infty} V.P. \int_{-\infty}^{\infty} dk \frac{(u_m u_k)^2}{\varepsilon - \varepsilon_{mk\sigma}},$$
(7)
$$g_\sigma(\varepsilon) = \sum_{m=-\infty}^{\infty} \int_{-\infty}^{\infty} dk (u_m u_k)^2 \delta(\varepsilon - \varepsilon_{mk\sigma}),$$

we obtain the I.M. Lifshitz dispersion equation for the local and resonance energy levels of electrons on the tube:

$$1 - \frac{u_0}{4\pi^2 a}[F_\sigma(\varepsilon) - i\pi g_\sigma(\varepsilon)] = 0.$$

The coupling constant $u_0$ has a dimension $erg \cdot cm$. The density of electron states with spectrum (5) is [52]

$$\nu(\varepsilon) = \frac{L}{\pi}\sqrt{2m_*} \sum_{m\sigma} \frac{\Theta(\varepsilon - \varepsilon_{m\sigma})}{\sqrt{\varepsilon - \varepsilon_{m\sigma}}},$$
(8)

where

$$\varepsilon_{m\sigma} = \varepsilon_0 \left(m + \frac{\Phi}{\Phi_0}\right)^2 + \sigma\mu B,$$

$\Theta$ is the Heaviside function, $L$ is the nanotube length ($L \to \infty$). It can be seen from Eq. (5) and Eq. (8) that the energy spectrum of an electron on a impurity-free tube consists of a set of one-dimensional subbands, the boundaries of which $\varepsilon_{m\sigma}$ are not equidistant. The density of states reaches a minimum at the "ceiling" of each subzone. These are the energies to look for the resonance levels split off downward by the impurity attraction from the following subband. The only local level in the potential considered here splits off from the lower boundary $\varepsilon_\sigma = \sigma\mu B$ of the continuous spectrum into the energy interval $\varepsilon < \varepsilon_\sigma$.

## 3. Calculation
### 3.1. Local level of electrons

In the energy range $\varepsilon < \varepsilon_\sigma$, the function $g_\sigma$ is equal to zero, the dispersion equation for the local level $\varepsilon = \varepsilon_l$ has the form

$$1 = -\upsilon_0 \frac{2m_*}{\pi a} \sum_m \exp\left(-\frac{m^2\eta}{2}\right) \int_0^\infty dk \frac{\exp\left(-\frac{k^2 b^2}{2}\right)}{k_{m\sigma}^2 + k^2},$$
(9)

where $k_{m\sigma} = \sqrt{2m_*(\varepsilon_{m\sigma} - \varepsilon)}$, $\upsilon_0 = u_0 b$ is the coupling constant with dimension $erg \cdot cm^2$. The integral entering into Eq. (9) is known [53]. As a result, we obtained

$$1 = -\frac{m_* \upsilon_0 \sqrt{\eta}}{a} \sum_m \frac{1}{k_{m\sigma}} \exp\left(-\frac{m^2\eta}{2}\right) \exp\left(\frac{k_{m\sigma}^2 b^2}{2}\right) \text{erfc}\left(\frac{k_{m\sigma} b}{\sqrt{2}}\right),$$
(10)

where $\text{erfc}(x)$ is the additional integral of probability [53]. Eq. (10) shows that its solution exists for $\upsilon_0 < 0$ only, which corresponds to the attraction of an electron to the impurity atom.

It is convenient to transform the right side of Eq. (10) using the Poisson formula [54]:

$$\sum_{m=-\infty}^{\infty} f(m) = \sum_{l=-\infty}^{\infty} \int_{-\infty}^{\infty} dx\, f(x)\exp(2\pi i l x). \tag{11}$$

Then the dispersion equation (10) becomes as follows:

$$1 = 2m_* |\upsilon_0| \sqrt{\eta} \left[ \int_0^{\infty} dx \frac{e^{-\frac{\eta x^2}{2}}}{\sqrt{x^2 + x_\sigma^2}} + 2\sum_{l=1}^{\infty} \cos\left(2\pi l \frac{\Phi}{\Phi_0}\right) \int_0^{\infty} dx \frac{e^{-\frac{\eta x^2}{2}}}{\sqrt{x^2 + x_\sigma^2}} \cos(2\pi l x) \right], \tag{12}$$

where

$$x_\sigma = \left(\frac{\varepsilon_\sigma - \varepsilon}{\varepsilon_0}\right)^{1/2}.$$

In Eq. (12), without breaking the convergence, the passage to the limit is made $b \to 0$, but $\eta \neq 0$. After changing the variable of integration $\eta x^2 / 2 = y$ in the first term in square brackets, we obtain the tabular integral [55]:

$$\int_0^{+\infty} \frac{e^{-y} dy}{\sqrt{y(y+z)}} = e^{z/2} K_0\left(\frac{z}{2}\right),$$

where $K_0$ is the Macdonald function [56],

$$z = \frac{\eta}{2}\left(\frac{\Delta}{\varepsilon_0}\right)^2, \qquad \Delta = \varepsilon_\sigma - \varepsilon_l$$

is the distance from the local level $\varepsilon_l$ to the boundary of the continuous spectrum. The appearance of the quantity $\Delta/\varepsilon_0$ makes it possible to speak of strong $(\Delta \gg \varepsilon_0)$ and weak $(\Delta \ll \varepsilon_0)$ localizations of the electron. In the case of strong localization, the second integral in square brackets in Eq. (12) is equal to [57]

$$\frac{1}{x_\sigma} \int_0^{\infty} dx\, e^{-\frac{\eta x^2}{2}} \cos 2\pi l x = \frac{1}{x_\sigma} \sqrt{\frac{\pi}{2\eta}} \exp\left(-\frac{2\pi^2 l^2}{\eta}\right).$$

Given the formula

$$K_0(x) \approx \sqrt{\frac{\pi}{2x}} e^{-x} \quad (x \to +\infty)$$

and the integrals presented here, from Eq. (12) in the case of strong localization we obtain

$$\Delta = \frac{\pi m_* \upsilon_0^2}{a^2} \left[1 + 4\sum_{l=1}^{\infty} \cos\left(2\pi l \frac{\Phi}{\Phi_0}\right) \exp\left(-\frac{2\pi^2 l^2}{\eta}\right)\right]. \tag{13}$$

The series included here is expressed in terms of the Jacobi elliptic theta function $\Theta_3$ [55, 57]:

$$\Delta = \frac{\pi m_* \upsilon_0^2}{a^2}\left[2\Theta_3\left(\frac{\Phi}{\Phi_0}, \frac{2}{\eta}\right) - 1\right]. \tag{14}$$

The quantity (13) undergoes Aharonov-Bohm oscillations with a change of the magnetic flux. The oscillation period is $\Phi_0$. The rapid decrease in the amplitude of the oscillations with an increase in the harmonic number $l$ suggests us that the magnitude of the splitting off (13) is approximately equal to

$$\Delta = \frac{\pi m_* \upsilon_0^2}{a^2}. \tag{15}$$

In the weak localization mode the first term in square brackets in Eq. (12) dominates:

$$1 = m_* |\upsilon_0| \sqrt{\eta} \exp\left(\frac{\Delta \eta}{4\varepsilon_0}\right) K_0\left(\frac{\Delta \eta}{4\varepsilon_0}\right). \tag{16}$$

Taking into account the relation $K_0(x) \approx -\ln x \ (x \to 0)$ from Eq. (16) we obtain

$$\Delta = \frac{2}{m_* a^2 \eta} \exp\left(-\frac{1}{m_* |\upsilon_0| \sqrt{\eta}}\right). \tag{17}$$

Up to numerical factors this expression coincides with the expression given in the textbook [58], obtained by another method. This indicates the importance of the need to smear the impurity potential in the azimuthal direction.

### 3.2. Resonant energy levels of electrons
### 3.2.1. Positions of the resonance levels

In Subsection 3.1, calculating sums and integrals, there were no restrictions on the limits of summation and integration, except for the requirement of convergence. In the case of resonances, the situation is different. They should be sought below the boundaries of the subzones from which they are split off by an impurity atom of attraction. This means that the denominator in Eq. (7) is still equal to

$$\varepsilon - \varepsilon_{mk\sigma} = -\frac{1}{2m_*}\left(k^2 + k_{m\sigma}^2\right),$$

where $k_{m\sigma} = \sqrt{2m_*(\varepsilon_{m\sigma} - \varepsilon)}$, but it is necessary to sum up over those $m$, for which $\varepsilon < \varepsilon_{m\sigma}$, from some minimum value $m$, at which $\varepsilon_{m\sigma} = \varepsilon$.

Using the Poisson formula (11), we represent the equation for the positions of the resonance levels in the form

$$1 = -\frac{m_* \upsilon_0 \sqrt{\eta}}{a} \sum_{l=-\infty}^{\infty} \int_{-x_\sigma}^{x_\sigma} dx \frac{1}{k_{x\sigma}} \exp\left(-\frac{\eta x^2}{2}\right) \exp\left(\frac{b^2 k_{x\sigma}^2}{2}\right) \mathrm{erfc}\left(\frac{bk_{x\sigma}}{\sqrt{2}}\right) e^{2\pi i l x}, \tag{18}$$

where

$$k_{x\sigma} = \sqrt{2m_* \varepsilon_0}\left[\left(x + \frac{\Phi}{\Phi_0}\right)^2 - x_\sigma^2\right]^{1/2}, \qquad x_\sigma = \left(\frac{\varepsilon_\sigma - \varepsilon}{\varepsilon_0}\right)^{1/2}.$$

After shifting the integration variable $x + \Phi/\Phi_0 \to x$ and performing transformations similar to those performed in Subsection 3.1, Eq. (18) has the form:

$$1 = -2m_* \upsilon_0 \sqrt{\eta} \left[ \int_{x_\sigma}^{\infty} dx \frac{e^{-\frac{\eta x^2}{2}}}{\sqrt{x^2 - x_\sigma^2}} + 2\sum_{l=1}^{\infty} \cos\left(2\pi l \frac{\Phi}{\Phi_0}\right) \int_{x_\sigma}^{\infty} dx \frac{e^{-\frac{\eta x^2}{2}}}{\sqrt{x^2 - x_\sigma^2}} \cos(2\pi l x) \right]. \tag{19}$$

The first integral on the right-hand side of Eq. (19) is [55]:

$$\int_{x_\sigma}^{\infty} dx \frac{e^{-\frac{\eta x^2}{2}}}{\sqrt{x^2 - x_\sigma^2}} = \frac{1}{2} e^{-\frac{\eta x_\sigma^2}{4}} K_0\left(\frac{\eta x_\sigma^2}{4}\right).$$

Using this integral, we obtain the positions of the resonance levels (13), (15), (17) $\varepsilon_r = \varepsilon_{m\sigma} - \Delta$ in the cases of strong and weak localization. The second integral in Eq. (19) remains finite at $\eta = 0$. It is equal to [55]:

$$\int_{x_\sigma}^{\infty} dx \frac{\cos(2\pi l x)}{\sqrt{x^2 - x_\sigma^2}} = -\frac{\pi}{2} Y_0(2\pi l x_\sigma),$$

where $Y_0$ is the Neumann function [56]. Equation (19) takes the form

$$1 = m_* |\upsilon_0| \sqrt{\eta} \left\{ \exp\left[-\frac{\eta(\varepsilon - \varepsilon_\sigma)}{4\varepsilon_0}\right] K_0\left[-\frac{\eta(\varepsilon - \varepsilon_\sigma)}{4\varepsilon_0}\right] - 2\pi \sum_{l=1}^{\infty} \cos\left(2\pi l \frac{\Phi}{\Phi_0}\right) Y_0\left(2\pi l \sqrt{\frac{\varepsilon - \varepsilon_\sigma}{\varepsilon_0}}\right) \right\}. \tag{20}$$

For $l x_\sigma \gg 1$, the asymptotics of the Neumann function is [56]

$$Y_0(x) \approx \sqrt{\frac{2}{\pi x}} \sin\left(x - \frac{\pi}{4}\right).$$

### 3.2.2. The widths of the resonance levels

To estimate the widths of the resonance levels by the formula

$$\Gamma = \frac{\pi g(\varepsilon_r)}{|F'(\varepsilon_r)|} \tag{21}$$

(prime denotes the derivative with respect to the electron energy at the point $\varepsilon = \varepsilon_r$), it is necessary to calculate the function $g(\varepsilon)$ in Eq. (7).

The function $g_\sigma(\varepsilon)$ is related to the density of states (8) and to the function $F_\sigma(\varepsilon)$:

$$F_\sigma(\varepsilon) = V.P. \int_{-\infty}^{\infty} d\varepsilon' \frac{g_\sigma(\varepsilon')}{\varepsilon - \varepsilon'}.$$

The summation index $m$ in this function is constrained by the condition $\varepsilon > \varepsilon_{m\sigma}$. Using Poisson's formula and omitting calculations duplicating this Section, we restrict ourselves to the final result:

$$g_\sigma(\varepsilon) = 4\pi^2 m_* ab \sqrt{\eta} \left\{ \exp\left[-\frac{\eta}{4}\left(\frac{\varepsilon - \varepsilon_\sigma}{\varepsilon_0}\right)\right] I_0\left[\frac{\eta}{4}\left(\frac{\varepsilon - \varepsilon_\sigma}{\varepsilon_0}\right)\right] + 2\sum_{l=1}^{\infty} \cos\left(2\pi l \frac{\Phi}{\Phi_0}\right) J_0\left(2\pi l \sqrt{\frac{\varepsilon - \varepsilon_\sigma}{\varepsilon_0}}\right) \right\},$$

where $I_0$ and $J_0$ are the modified and ordinary Bessel functions [56].

Let us estimate the width $\Gamma_1$ of the sharpest resonance located near the upper boundary of the lower subband. In this case, the function $g_\sigma(\varepsilon)$ contains one term with $m = 0$ only, and the summation in the function $F_\sigma(\varepsilon)$ starts with $m = 1$. Keeping in function $F_\sigma$ one term with a root singularity at the upper boundary of the first subzone only, from Eq. (21) we obtain

$$\Gamma_1 = \frac{2}{\pi}\Delta\sqrt{\frac{\Delta}{\varepsilon_0}}, \qquad (22)$$

where $\Delta$ is the distance between the upper boundary of the first subzone and the resonant level. From Eq. (22) it is seen that $\Gamma_1 \ll \Delta$, if $\Delta \ll \varepsilon_0$. The same technique can be used to estimate the resonance widths in the higher subbands. They increase, since the density of states increases with the increasing subband number.

## 4. Results

The positions of the local energy levels of electrons are found in the cases of strong and weak localizations in the field of an attractive impurity. The binding energy of an electron experiences Aharonov-Bohm oscillations with a change in the magnetic flux.

The positions of the resonance levels are found. They also experience the Aharonov-Bohm oscillations. Oscillations that are not associated with the magnetic field are superimposed on them. The resonance width is estimated.

## 5. Discussion

In this paper, an impurity atom is approximated by a potential of a special type that allows an analytical solution to the problem. The potential proposed here is characterized by three parameters: the intensity and lengths of the potential in the azimuthal direction and in the direction along the nanotube axis. Thus, the model considered here is richer than the model with two parameters, which was previously used in the paper [41].

As can be seen from Eq. (8), the density of states has root singularities at the boundaries of the subbands $\varepsilon_{m\sigma}$. This character of the density of states is analogous to the density of states of electrons in a bulk sample in a quantizing magnetic field, which makes it similar to the one-dimensional motion of electrons. This means that the quasilocal (resonance) energy levels of electrons on the tube will have a minimum width if they are split off by an attractive impurity from the boundaries of the subbands down the energy axis to the region where the density of states in the subband is minimal. This also applies to magnetoimpurity states of electrons in metals and semiconductors under conditions when a weak impurity in the absence of a magnetic field cannot localize an electron [32–34].

It follows from the transcendental dispersion equation (20) that the positions of the resonance levels experience Aharonov-Bohm oscillations with a change in the magnetic flux. They also experience oscillations that are not associated with a magnetic field. The latter are due to the quantization of the energy of the circular motion of electrons and are similar to the de Haas-van Alphen oscillations. Resonance levels play the role of Landau levels on the tube only. Oscillations of the de Haas-van Alphen type should manifest themselves when studying observable quantities, for example, the heat capacity of a degenerate electron gas on a tube. They are due to the passage of resonance levels through the Fermi boundary with a change in the parameters of the problem.

# 6. Conclusion

The localization of electrons in the field of isolated nonmagnetic impurity atoms on the surface of a semiconductor or carbon nanotube with metallic conductivity in a magnetic field is considered. Attempts to solve this problem within the framework of the model of the $\delta$-potential of an impurity atom run into the divergence of the integral in the dispersion equation for impurity energy levels. This requires the use of methods of regularization of integrals and renormalizations of quantum field theory. This paper uses a different method to eliminate divergences. The proposed I.M. Lifshitz model of a Gaussian separable impurity potential is employed. This potential is "smeared" along the nanotube axis and in the azimuthal direction. Within the framework of this model, the positions of the local energy levels of electrons are found in the cases of strong and weak localizations in the field of an attractive impurity. The binding energy of an electron experiences Aharonov-Bohm oscillations with a change in the magnetic flux. The positions of the resonance levels are found. They also experience the Aharonov-Bohm oscillations. Oscillations that are not associated with the magnetic field are superimposed on them. They are due to the quantization of the energy of the circular motion of electrons. Resonance levels should manifest themselves in the study of thermodynamic and kinetic quantities when they cross the Fermi boundary with changing magnetic field and nanotube parameters. The resonance width is estimated. These effects can be also discovered by studying the optical properties of a GaAs/AlGaAs heterojunction bent into a cylinder with a two-dimensional electron gas doped with donor impurities.

## Acknowledgment

The author is very grateful to professor A. M. Ermolaev for fruitful discussion of the results of this paper.
The author would like to thank Dr. T. I. Rashba for help with preparation of the manuscript.